# Au-Gated SrTiO$_3$ Field-Effect Transistors with Large Electron Concentration and Current Modulation

Amit Verma[1*], Santosh Raghavan[2], Susanne Stemmer[2] and Debdeep Jena[1]

[1]Department of Electrical Engineering, University of Notre Dame, Notre Dame, Indiana 46556, U.S.A.
[2]Materials Department, University of California, Santa Barbara, California 93106, U.S.A.
*E-mail: averma@nd.edu

Abstract: We report the fabrication of low-leakage rectifying Pt and Au Schottky diodes and Au-gated metal-semiconductor field effect transistors (MESFETs) on n-type SrTiO$_3$ thin films grown by hybrid molecular beam epitaxy. In agreement with previous studies, we find that compared to Pt, Au provides a higher Schottky barrier height with SrTiO$_3$. As a result of the large dielectric constant of SrTiO$_3$ and the large Schottky barrier height of Au, the Au-gated MESFETs are able to modulate ~1.6 x 10$^{14}$ cm$^{-2}$ electron density, the highest modulation yet achieved using metal gates in any material system. These MESFETs modulate current densities up to ~68 mA/mm, ~20x times larger than the best demonstrated SrTiO$_3$ MESFETs. We also discuss the roles of the interfacial layer, and the field-dependent dielectric constant of SrTiO$_3$ in increasing the pinch off voltage of the MESFET.

Strong electron-electron interactions in complex oxide crystals give rise to many interesting emergent phenomena such as high-T$_c$ superconductivity, metal-insulator transitions, and colossal magnetoresistance [1-4]. Reversible control of such phenomena by changing the carrier concentration using a field-effect is a promising pathway for tuning electron correlations and for using these materials for practical applications [5-11]. Realizing such devices is however challenging because the carrier concentration modulation needed to change correlated electron behavior is of the order of ~10$^{14}$ cm$^{-2}$ and greater [5-10]. This concentration modulation is almost one order higher than the modulation typically achieved in traditional semiconductors, and therefore puts stringent requirements on the gate dielectric design. To modulate large carrier concentrations, the gate dielectric should have a large dielectric constant (large capacitance), low leakage and large breakdown field. Ionic liquids can satisfy all these conditions and have been used widely to modulate complex oxide properties [7-10]. However, ions have large mass and respond quite slowly to applied electric fields. As a result, transistors with ionic liquids as gate dielectrics can operate only at very low frequencies [12]. For practical applications, higher modulation frequencies are required and metal gates are therefore needed. To achieve large carrier concentration modulation using metal gates, we can use high-k oxides such as $SrTiO_3$ as the gate dielectric. In addition to having a high room temperature dielectric constant of ~300, $SrTiO_3$ also has a large bandgap (~3.2 eV) and therefore a large breakdown strength. Recently, electron concentration modulation of ~1.1 x$10^{14}$ cm$^{-2}$ of a two-dimensional electron gas at the $SrTiO_3$/$GdTiO_3$ interface was reported using $SrTiO_3$ as a gate dielectric with Pt as the gate metal [13].

In this work, using a higher Schottky-barrier height metal Au as gate in $SrTiO_3$ MESFETs, we improve the electron concentration modulation to ~1.6 x $10^{14}$ cm$^{-2}$, almost by ~50% compared to previous work [13]. In this effort we also improve the current density modulated in a $SrTiO_3$ MESFET by ~20x, more than an order higher than earlier reports [13]. On measuring the transistor characteristics of these FETs, we observe the pinch off voltage to be significantly larger than predicted by standard device physics. We ascribe this discrepancy to the presence of a low dielectric constant interfacial layer at the gate metal/$SrTiO_3$ interface. We also discuss the relative importance of the electric field-dependent dielectric constant of $SrTiO_3$ in increasing the pinch off voltage of the MESFET.

For fabricating the $SrTiO_3$ Schottky diodes and MESFETs, 160 nm thick $SrTiO_3$ thin films were grown on insulating (001) oriented $(LaAlO_3)_{0.3}(Sr_2AlTaO_6)_{0.7}$ (LSAT) substrates using hybrid molecular beam epitaxy (MBE). In this growth technique, Ti and O are provided using the organometallic precursor titanium tetra isopropoxide, and Sr is provided using an effusion cell. Additional O can be provided using an oxygen plasma source. However, for this study, oxygen plasma source was not used. The resultant oxygen deficient growth condition created vacancies, doping the sample with mobile electrons of concentration ~$10^{19}$ cm$^{-3}$. More details on the growth procedure and structural quality of grown epilayers have been reported elsewhere [14-16].

Pt and Au gated diodes and MESFETs were fabricated on two halves of a grown sample. For isolating the devices, mesa etching was first performed using an inductively coupled plasma-reactive ion etching (ICP-RIE) system. A $BCl_3$/Ar plasma (45/5 sccm, 5 mTorr, 1000 W ICP, 75 W RIE) was used for etching. Subsequently, Al/Ni/Au (40/40/100 nm) ohmic contact metal stacks were deposited using e-beam evaporation. A contact resistance of ~0.7 Ω-mm was extracted using transmission line measurements. An electron mobility of ~5.3 cm$^2$/V-sec and sheet electron concentration of ~1.28 x $10^{14}$ cm$^{-2}$ was obtained from Hall-effect measurements performed in a Van der Pauw geometry. This sheet electron concentration value is lower than the expected value of ~1.6 x $10^{14}$ cm$^{-2}$ because of surface depletion in the SrTiO$_3$ thin film [17]. For a comparative gate stack study, Pt/Au (40/100 nm) and 160 nm Au was deposited by e-beam evaporation on the two samples respectively. Prior to the Schottky metal deposition, Oxygen plasma treatment (20sccm, 333 mTorr, 16 W) of the SrTiO$_3$ surface was performed in a RIE system to reduce the gate leakage and improve rectification [18]. No significant change in electron mobility and sheet electron concentration was observed in Hall-effect measurements done after the Oxygen plasma treatment.

For measuring the device characteristics, a Cascade probe station was used with a Keithley 4200 semiconductor characterization system. I-V characteristics of typical fabricated Pt and Au Schottky diodes are shown in Fig. 1a. The characteristics shown are from 15 μm radius circular Schottky diodes. Clearly, both Pt and Au diodes exhibit rectifying characteristics. From the forward bias characteristics, the barrier height with Pt was found to be ~0.87 eV with an ideality factor ~ 1.46 and that of Au to be ~1.01 eV with an ideality factor ~1.57. In agreement with earlier reports, Au exhibits a larger Schottky barrier height with SrTiO$_3$ as compared to Pt, suggesting the greater charge modulation potential of Au gates [19]. Both Schottky barrier height values in our devices are about ~0.1 eV higher compared to the previous report [19]. The reasons for this difference can be the lower doping in our samples that leads to lower image-force lowering, and a different surface preparation recipe before the Schottky metal deposition. Capacitance-Voltage (C-V) measurements (frequency 100 kHz, signal amplitude 30 mV) of Pt and Au diodes are shown in Fig. 1b and Fig. 1c respectively. As a result of the large dielectric constant of SrTiO$_3$, high capacitance values of ~2-3 μF/cm$^2$ are easily achievable in these diodes, suggesting the promise of SrTiO$_3$ as a gate dielectric for modulating large carrier concentrations.

The drain current-voltage $I_{ds}$-$V_{ds}$ characteristics of a 1.85μm gate length Au-gated MESFET are shown in Fig. 2a. This figure shows that we are able to modulate 54 mA/mm current density or equivalently ~1.28 x $10^{14}$ cm$^{-2}$ of the electron concentration (as measured using Hall-effect) in SrTiO$_3$. This current density modulation is ~20x times larger than the earlier reported values in SrTiO$_3$ MESFET [13]. During measurement of $V_{gs}$-$I_{ds}$ transfer characteristics of the MESFET shown in Fig.2b, a drain voltage of 10V has been applied. This positive drain bias allows applying forward bias to the gate without causing gate leakage. Under this condition, we are able to modulate up to 68.6 mA/mm of current density or an equivalent carrier density of ~1.62 x $10^{14}$ cm$^{-2}$. This carrier density modulation is an improvement of almost ~50% over the previous reported value [13]. Also, the log scale $V_{gs}$-$I_{ds}$ transfer characteristics in Fig. 2b show that we are able to achieve about 3 orders of ON/OFF current ratio and that the gate leakage current $I_g$ is significantly lower than $I_{ds}$ in the ON state. On reversing the scan direction of the gate voltage, we observe some hysteresis in $I_g$ which is typical of metal/SrTiO$_3$ Schottky junctions. However, this hysteresis does not affect the modulation achieved in MESFET significantly. We are able to pinch off the MESFETs in both negative to positive and positive to negative scans of the gate terminal with more than two orders of ON/OFF current ratio.

From a linear fit to the $\sqrt{I_{ds}}$ vs $V_{gs}$ characteristics of the MESFET, we extract a threshold voltage of -17.2 V. This corresponds to a measured pinch off voltage of $V_{Pm}$ = 17.2 + 1.01 V ~18.2 V, where 1.01 V is the built-in bias ($V_{bi}$) due to the Au Schottky barrier. The expected value of the pinch off voltage can be calculated as [20],

$$V_{P1} = \frac{qN_D d^2}{2\varepsilon_0 \varepsilon_r} \tag{1}$$

where, $q$ is the electron charge, $N_D = 10^{19}$ cm$^{-3}$ is the doping density, $d = 160$ nm is the doped film thickness, $\varepsilon_0$ is the vacuum permittivity, and $\varepsilon_r = 300$ is the low field dielectric constant of SrTiO$_3$. The expected pinch off voltage $V_{P1}$ as a function of doped SrTiO$_3$ film thickness is plotted in Fig. 3b. For a film thickness of 160 nm, the calculated $V_{P1}$ is ~7.7 V. Clearly, the measured pinch-off voltage is much larger compared to the expected value. Two factors can contribute to this pinch off voltage increase: the field-dependent dielectric constant of SrTiO$_3$, and the presence of an interfacial low dielectric constant layer (so called dielectric dead layer) at the metal/SrTiO$_3$ interface.

The dielectric constant of SrTiO$_3$ decreases with the increase in applied electric field according to the empirical formula, $\varepsilon_r(E) = b/\sqrt{a + E^2}$, where $a = 7.03 \times 10^{15}$ V$^2$/m$^2$, $b = 2.51 \times 10^{10}$ V/m are constants and $E$ is the local electric field [21, 22]. The pinch off voltage after including this effect is [23],

$$V_{P2} = \frac{\sqrt{ab}\varepsilon_0}{qN_D}\left[\cosh\left(\frac{qN_D d}{b\varepsilon_0}\right) - 1\right] \tag{2}$$

$V_{P2}$ as a function of the doped SrTiO$_3$ film thickness is also shown in Fig. 3b. For 160 nm SrTiO$_3$ film thickness, $V_{P2}$ ~8.6 V. This pinch off voltage is still quite small compared to the measured pinch off voltage, suggesting the existence of some other factor responsible for increasing the pinch off voltage.

It is well known that the capacitance of thin film capacitors of high-k materials is smaller than expected and is usually ascribed to the presence of a thin interfacial layer with much lower dielectric constant compared to the bulk [13, 21, 24-26]. Gauss's law requires the continuity of the displacement vector across the interface; because of the low dielectric constant of the interfacial layer, a large fraction of applied voltage drops across it (Fig. 4). If the interfacial layer thickness is small compared to the doped film thickness, the pinch off voltage of a MESFET with an interfacial layer of capacitance $C_i$ is calculated to be,

$$V_{P3} \approx \frac{qN_D d^2}{2\varepsilon_0 \varepsilon_r} + \frac{qN_D d}{C_i}, \tag{3}$$

where the first term is the voltage drop across the doped SrTiO$_3$ thin film and the second term is the voltage drop across the interfacial capacitance. To calculate $V_{P3}$ for our MESFET device structure, we need an estimate of $C_i$. Since, this interfacial capacitance acts in series with the depletion capacitance, we can estimate its value from the reduction in measured capacitance of Au Schottky diodes compared to the expected capacitance value. The expected Schottky diode depletion capacitance is given as, $C_{d,ex}(V) = \sqrt{q\varepsilon_0 \varepsilon_r N_D / 2V}$, where $V$ is the total voltage drop across the Schottky depletion region and for zero applied bias case, $V = V_{bi}$ [20]. For $\varepsilon_r$ = 300, $N_D$ = 10$^{19}$ cm$^{-3}$, and $V$ = 1.01 V, the expected zero bias capacitance for Au Schottky diode is ~ 4.59 µF/cm$^2$. The measured zero bias capacitance value in Au Schottky diode is smaller ~2.3 µF/cm$^2$, about half the expected value. The net capacitance is the series combination of $C_i$ and the depletion capacitance. However, the zero bias depletion capacitance value would be different from the expected value of 4.59 µF/cm$^2$, because part of the built-in voltage drops across $C_i$. The modified depletion thickness in the presence of an interfacial capacitance layer can be expressed in terms of the total voltage drop across the device by inverting eqn. 3,

$$d(V) \approx -\frac{\varepsilon_0 \varepsilon_r}{C_i} + \sqrt{\left(\frac{\varepsilon_0 \varepsilon_r}{C_i}\right)^2 + \frac{2\varepsilon_0 \varepsilon_r V}{qN_D}} \tag{4}$$

and the corresponding depletion capacitance would be given as $C_d(V) = \varepsilon_0 \varepsilon_r / d(V)$. For zero applied bias case $V = 1.01$ V. Now, we have expressed $C_d(V)$ in terms of the interfacial capacitance $C_i$. Solving, $C_d(V) C_i / (C_d(V) + C_i) = 2.3$ µF/cm² for $C_i$, we find the interfacial capacitance to be $C_i \sim 2.66$ µF/cm². The match between the measured capacitance and the series model $C_d(V) C_i / (C_d(V) + C_i)$ for the Au Schottky diode is shown in Fig. 3a. Now, using eqn. 3, we can calculate the pinch off voltage $V_{P3}$ when an interfacial layer with capacitance $C_i \sim 2.66$ µF/cm² is present. $V_{P3}$ as a function of doped SrTiO₃ film thickness is shown in Fig. 3b. For 160 nm film thickness, $V_{P3} \sim 17.3$ V, quite close to the measured value of 18.2 V. This agreement with the measured value of the capacitance (Fig.3a) and pinch off voltage (Fig.3b) suggests that the interfacial layer is the main reason for the increase in the pinch off voltage of the MESFET, while the field dependent dielectric constant of SrTiO₃ plays only a minor role in the on-state characteristics.

Recently, it has been shown that these interfacial layers are a result of surface contamination that is insufficiently removed during room temperature gate metal processing [25]. These layers also affect the effective Schottky barrier height of the metal/SrTiO₃ interface and can therefore impact the reverse leakage currents [25]. It is therefore anticipated that with better processed and optimized metal electrodes, higher Schottky barrier heights and large gate capacitances can be achieved simultaneously.

To summarize this work, we have demonstrated record electron concentration modulation of $\sim 1.6 \times 10^{14}$ cm⁻² and record current modulation of ~68.6 mA/mm in SrTiO₃ MESFETs using a higher Schottky barrier metal (Au) gate. This carrier density modulation is highest yet achieved using metal gates in any semiconductor device and the MESFET current density is a ~20x improvement compared to earlier SrTiO₃ MESFET reports. We have found that the pinch off voltage of SrTiO₃ MESFETs increases due to the presence of interfacial dielectric dead layer at the gate metal/SrTiO₃ interface. We hope this work will help enable further improvements in large carrier concentration modulation in complex oxide devices, ultimately leading to reversible manipulation of emergent phenomena in these materials.

This work was supported by the Extreme Electron Concentration Devices (EXEDE) MURI program of the Office of Naval Research (ONR) through grant No.N00014-12-1-0976.

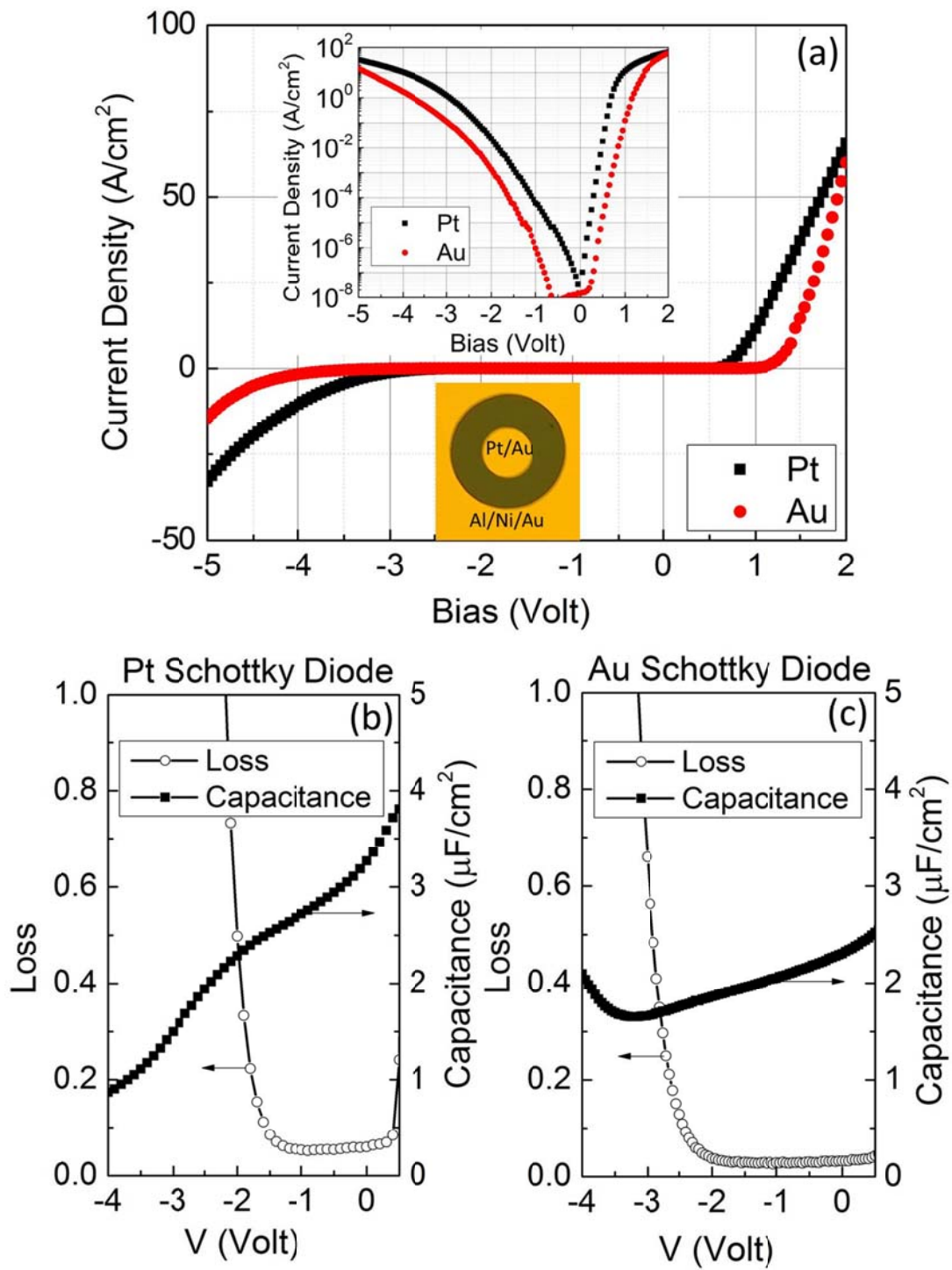

**Fig 1**. a) Measured I-V characteristics of Pt and Au circular Schottky diodes with 15 μm radius (Inset: Optical image of the Pt Schottky diode), (b, c) Measured C-V characteristics (100 kHz, 30mV) of (b) Pt and (c) Au Schottky diodes.

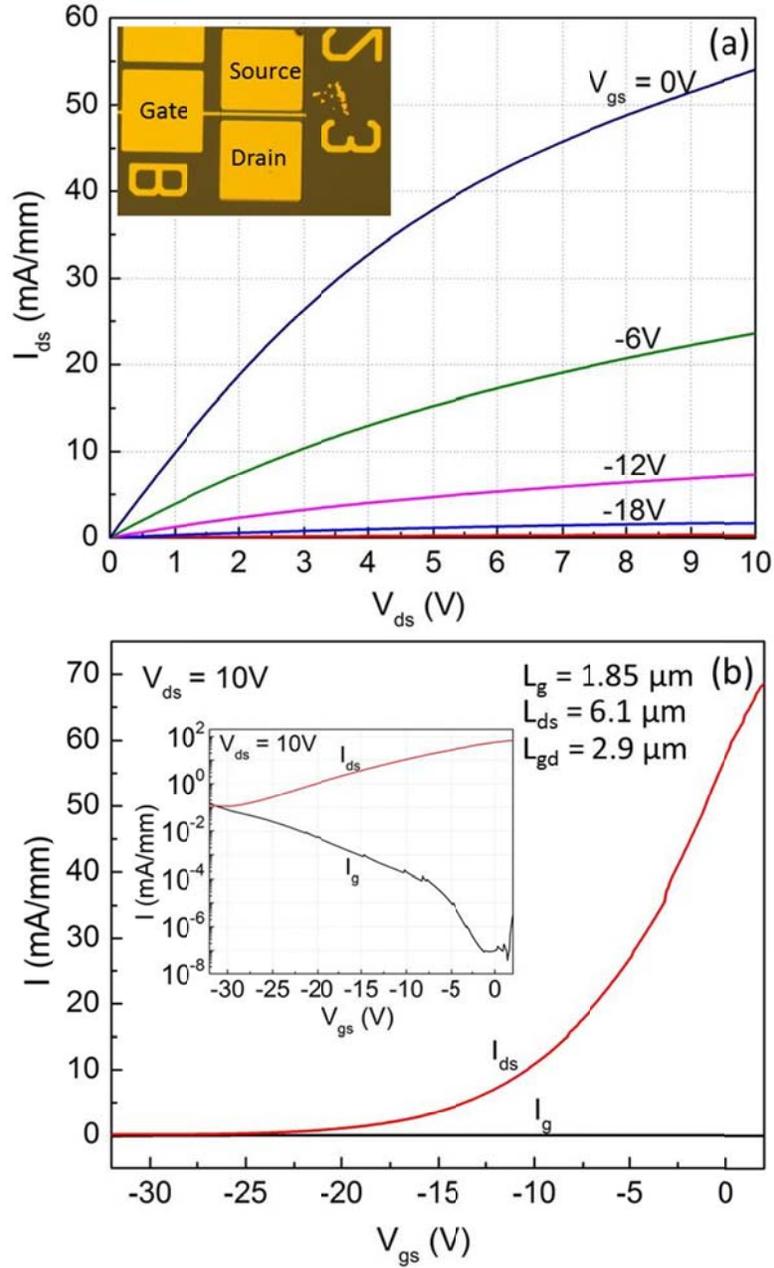

**Fig 2.**(a) $I_{ds}$-$V_{ds}$ characteristics of the Au-gated 160nm SrTiO$_3$ MESFET showing current modulation of 54 mA/mm or equivalently an electron concentration modulation of ~1.28 x 10$^{14}$ cm$^{-2}$, (Inset: Optical image of the MESFET device), (b) $I_{ds}$-$V_{gs}$ characteristics of the device measured at a drain bias of 10V demonstrating a current modulation of ~68.6 mA/mm and an ON/OFF ratio of 3 orders, equivalent electron concentration modulation is ~1.62 x 10$^{14}$ cm$^{-2}$. Measured device dimensions (Gate length $L_g$, Drain-Source separation $L_{ds}$, Gate-Drain separation $L_{gd}$) are also given.

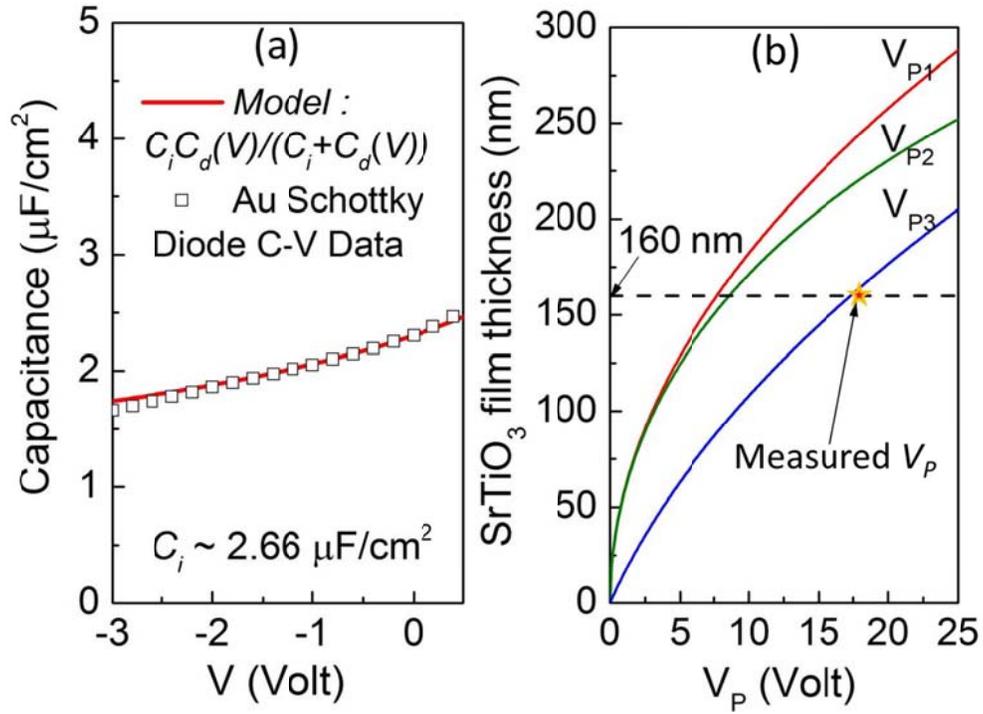

**Fig 3**. (a) Match between measured Au Schottky diode C-V data and a series capacitance model (interfacial capacitance $C_i$ (~2.66 µF/cm²) in series with the depletion capacitance $C_d(V)$), (b) Pinch off voltage plotted as a function of doped ($N_d$ ~$10^{19}$ cm⁻³) SrTiO₃ film thickness, different curves correspond to pinch off voltage in a: Normal MESFET ($V_{P1}$), MESFET with field dependent dielectric constant ($V_{P2}$), MESFET with interfacial dielectric dead layer ($V_{P3}$).

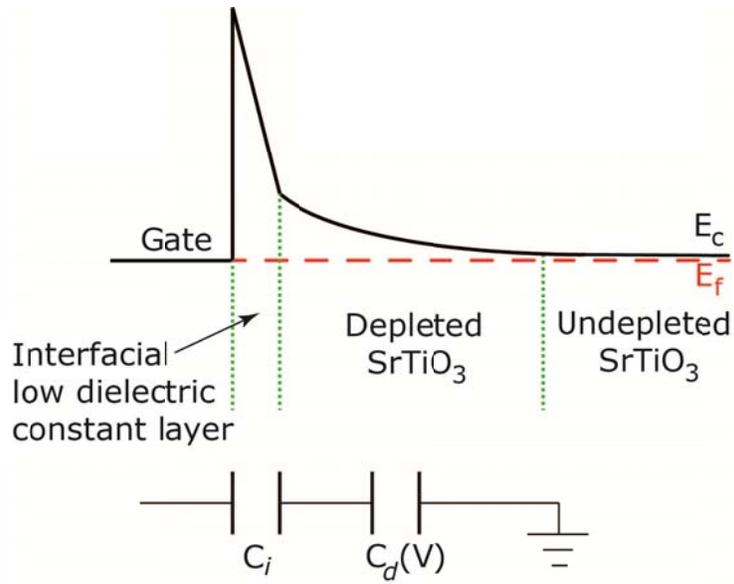

**Fig 4**. Schematic energy band diagram of a Schottky metal/SrTiO$_3$ junction showing the large voltage drop across the interfacial low dielectric constant layer and the effective capacitance model.